\documentclass[runningheads,a4paper]{ctextemp_llncs}

\usepackage{amssymb}
\usepackage{graphicx}

\usepackage{url}
\usepackage{multirow}
\usepackage{algorithm}
\usepackage{algorithmic}
\usepackage{subfigure}
\usepackage{balance}
\usepackage{amsmath}
\usepackage{xcolor}
\usepackage{amsfonts}
\usepackage{amssymb}
\usepackage{graphicx}

\usepackage{url}
\urldef{\mailsa}\path|{zhuyadong}@software.ict.ac.cn|
\urldef{\mailsb}\path|{lanyanyan, guojiafeng, cxq}@ict.ac.cn|

\begin{document}

\mainmatter  

\title{Topic-focused Dynamic Information Filtering in Social Media}

\titlerunning{Dynamic Information Filtering}

%
%

%
%

\author{Yadong Zhu
\and Yanyan Lan\and Jiafeng Guo\and Xueqi Cheng}

\institute{Institute of Computing Technology, Chinese Academy of Sciences, \\
Beijing 100190, China \\
\mailsa\\
\mailsb
}



%
\maketitle

\begin{abstract}
With the quick development of online social media such as twitter or sina weibo in china,
many users usually track hot topics to satisfy their desired information need.
For a hot topic, new opinions or ideas will be continuously produced in the form of \textit{online data stream}.
In this scenario, how to effectively filter and display information for a certain topic dynamically, will be a critical problem. We call the problem as \textit{Topic-focused Dynamic Information Filtering} (denoted as TDIF for short) in social media.
In this paper, we start open discussions on such application problems.
We first analyze the properties of the TDIF problem, which usually contains several typical requirements: relevance, diversity, recency and confidence. Recency means that users want to follow the recent opinions or news. Additionally, the confidence of information must be taken into consideration. How to balance these factors properly in online data stream is very important and challenging.
We propose a dynamic preservation strategy on the basis of an existing feature-based utility function, to solve the TDIF problem. Additionally, we propose new dynamic diversity measures, to get a more reasonable evaluation for such application problems. Extensive exploratory experiments have been conducted on TREC public twitter dataset, and the experimental results validate the effectiveness of our approach.
\end{abstract}

\section{Introduction}

The development of new social media such as twitter or sina weibo \footnote{http://weibo.com} accelerates the spread
of online information. In the social media, new information will be continuously produced in the form of online data stream, and how to retrieval useful information effectively will be very challenging. Specially, for a hot topic, how to filter and display relevant information dynamically will be a critical problem, which can be called as \textit{Topic-focused Dynamic Information Filtering} in social media.

The TDIF problem has three typical requirements: \textit{relevance}, \textit{diversity} and \textit{recency}.
The relevance requires the tweet information must be relevant to the topic. The diversity requires that
corresponding tweet information can describe the topic from different aspects with little redundancy.
Recency means that users want to follow the recent opinions or news quickly.
Additionally, the human factor also affects the confidence of the tweet information.
For example, the tweet information released by users with ``\textbf{V}'' authentication in sina weibo are usually with high confidence.
Therefore, how to balance these critical factors becomes a new challenging problem.

In fact, little prior research work has been done to tackle the TDIF problem. Most existing work only focuses one or two
factors in information retrieval, such as pure relevance \cite{Xia:20086,liu:LTR}, or pure diversity \cite{Yue:2008},
or relevance combing with diversity \cite{MMR,Agrawal:2009,zhu:2014:rRLTR}.
Even in the industry field, such problem has also been not solved well. They usually only consider relevance, but can not capture diversity or recency, such as sina weibo in china.

In this paper, We utilize the relational learning-to-rank model (R-LTR for short) \cite{zhu:2014:rRLTR} as utility function, and ccombine with the dynamic preservation scheme based on time periodic windows, to solve the TDIF problem.
R-LTR model is the state-of-the-art diverse ranking method, which models the diversity relations among documents in the ranking process, besides the content information of individual documents. It is a flexible feature-based ranking model with good adaptation to different application scenario.
Although R-LTR model can tackle relevance and diversity well, it is limited in the \textit{static} dataset. What's more, R-LTR model is with time complexity of $O(n*k)$, $n$ means the number of all the candidate objects, and $k$ indicates the number of desired results returned. Obviously, its efficiency can hardly satisfy the scenario of online data stream.

Therefore, we propose the dynamic preservation scheme based on the R-LTR model for proper solution.
Specifically, we segment the data stream into disjoint periods with time length $T$ (segmentation granularity can be days or hours depending on detailed requirements). For each new time window, we preserve the top-$(k-m)$ most relevant results previously, then utilize the R-LTR ranking function to select new $m$ relevant results, and finally display all the $k$ results in chronological order. Here the parameter $m$ can flexibly control the ``staleness'' of the returned results depending on the requirements of scenario.

Additionally, due to the new properties of TDIF application problem, we also propose new dynamic diversity evaluation measures to get a more reasonable evaluation. In these new measures, we introduce the \textit{recency} factor and \textit{confidence} factor into existing popular diversity evaluation measures (i.e.~$ERR\mbox{-}IA$\cite{Agrawal:2009,ERR}, $\alpha\mbox{-}NDCG$\cite{Clarke:2008} and $NRBP$\cite{NRBP}.).
Then we get a series of dynamic diversity evaluation measures: $d\mbox{-}ERR$, $d\mbox{-}NDCG$ and $d\mbox{-}NRBP$.

We conduct extensive evaluations on public TREC twitter dataset, and the experimental results show that our approach can achieve promising performance on both traditional diversity measures and new dynamic diversity measures.
Meanwhile, our approach is also with high processing efficiency.

The rest of the paper is organized as follows. Section 2 introduces our proposed approach for TDIF problem. Section 3
introduces the new dynamic diversity evaluation measures. Section 4 presents the experimental results. Section 5 describes related work and Section 6 concludes the paper.

\section{Our Approach}

As described before, the TDIF problem in social media has several typical requirements: relevance, diversity, recency and confidence.
Therefore, the basic motivation of our approach is how to effectively capture and balance these typical requirements.
In this section, we will describe our strategy for dynamic information filtering, which mainly contains two parts.
The first part is the chosen of basis utility function. The second part is the design of dynamic strategy that can take recency into consideration effectively.

\subsection{Utility Function}

The R-LTR model can effectively solve the diverse ranking problem in static dataset scenario, which models both relevance and diversity properly. As described in the literature \cite{zhu:2014:rRLTR}, the score of a candidate document contains two parts: relevance score based on content information of individual documents, and diversity score based on the relationship between the current document and those previously selected.
We use $X$ denotes all the candidate documents, $S$ denotes previously selected documents, and $X\backslash S$ denotes the remanent documents. The score function can be formalized as follows.
\begin{equation}
\label{eq:comb}
f_S(x_i,R_i)=\omega_r^T\textbf{x}_i+\omega_d^T h_S(R_i),\forall x_i\in X\backslash S
\end{equation}
where $\textbf{x}_i$ denotes the relevance feature vector of the candidate document $x_i$,
$R_i$ stands for the \textit{matrix} of relationships between document $x_i$ and other selected documents, with each $R_{ij}$ stands for the diversity feature vector between document $x_i$ and $x_j$, represented by the \textit{feature vector} of $(R_{ij1},\cdots,R_{ijl})$, $x_j \in S$, and $R_{ijk}$ stands for the $k$-th \textit{diversity feature} between documents $x_i$ and $x_j$.
$h_S(R_i)$ stands for the \textit{relational function} on $R_i$, $\omega_r^T$ and $\omega_d^T$ stands for the corresponding relevance and diversity weight vector.

The relational function $h_S(R_{i})$ denotes the way of representing the diversity relationship between the current document $x_i$ and the previously selected documents in $S$. It can be defined in three ways: Minimal, Average and Maximal. Here we choose the Minimal way, defined as follows.
\begin{displaymath}
h_S(R_i)=(\min_{x_j\in S} R_{ij1},\cdots,\min_{x_j\in S} R_{ijl}).
\end{displaymath}

As described above, the R-LTR is a flexible feature-based ranking function, which has good adaptation to social media scenario and can be chosen as our basis utility function.
Comparing with other heuristic definitions of utility function such as ``Max-Sum'' or ``Max-Min'' \cite{Gollapudi:2009,DrosouP09,Minack:2011}, we can obtain a more reasonable basis utility function by supervised learning.
When in real application, we need define and utilize specific relevance and diversity features close related to social media scenario.

\subsubsection{Relevance Feature Vector $\textbf{x}_i$.}

For relevance feature vector, we first utilize traditional learning-to-rank relevance features, shown as follows.
\begin{itemize}
\item Weighting Features. The typical weighting models include TF-IDF, BM25 and language model. For language model, we use query-likelihood language model with Dirichlet prior.

\item Term Dependency Features. We also employ the classic term dependency features such as MRF \cite{Metzler:2005:MRF}, to enhance relevance. The MRF has two types of values: ordered phrase and unordered phrase, so the total feature number is 2.
\end{itemize}

Additionally, we utilize some specific features in twitter, shown as follows.
\begin{itemize}
\item Recency. We take the time factor into consideration, and prefer more recent tweet information.
\item UserRank. The importance of a certain user account, which can capture the confidence of information. It can be simply obtained via the followers of account.
\item Retweet Number. If a tweet is retweet many times, it is usually with high importance.
\end{itemize}
This two types of features can be obtained via API\footnote{https://github.com/lintool/twitter-tools/wiki/TREC-2013-API-Specifications} provided by TREC.

\subsubsection{Diversity Feature Vector $R_{ij}$}
For diversity features, we utilize typical semantic diversity features shown as follows.

\textbf{Cosine Diversity.} The cosine diversity between two tweets is calculated based on their weighted term vector representations, and define the feature as follows.
\begin{displaymath}
R_{ij1}=1-\frac{\textbf{s}_i \cdot \textbf{s}_j}{\Vert {\textbf{s}_i} \Vert \Vert \textbf{s}_j \Vert}
\end{displaymath}
where $\textbf{s}_i$, $\textbf{s}_j$ are the weighted term vectors of tweets based on $tf*idf$, and $tf$ denotes the term frequencies, $idf$ denotes inverse term frequencies.

\textbf{Jaccard Diversity.} The Jaccard diversity between two tweets measures the ratio of overlapped terms, and is defined as follows.
\begin{displaymath}
R_{ij2}=1-\frac{\vert S_i \cap S_j \vert}{\vert S_i \cup S_j \vert}
\end{displaymath}
where $S_i$, $S_j$ are the term vectors of tweets.

\textbf{Subtopic Diversity.} Different tweets may associate with different aspects of the given topic. We use Probabilistic Latent Semantic Analysis (PLSA) \cite{Hofmann:1999} to model implicit subtopics distribution of candidate tweets. Then we can define a kind of subtopic diversity feature based on the KL distance, as follows.
\begin{displaymath}
R_{ij3}= \sum_{z_i \in Z}{P(z_i|S_i)log\frac{P(z_i|S_i)}{P(z_i|S_j)}}
\end{displaymath}
\begin{displaymath}
P(z_i|S_i)={\frac{1}{|S_i|}}{\sum_{w_j \in S_i}{P(z_i|S_i,w_j)}}
\end{displaymath}
where $P(z_i|S_i,w_j)$ is calculated and saved in the E-step of the EM procedure.

Based on these diversity features, we can obtain the diversity feature vector $R_{ij}=(R_{ij1},R_{ij2},R_{ij3})$.
Please note that here we only list some representative diversity features used in our work, other useful diversity features can be easily adopted into the utility function.

\subsection{Dynamic Preservation Scheme based on Periodic Windows}

\begin{figure}[!htb]
\begin{center}
  \includegraphics[width=2.8in]{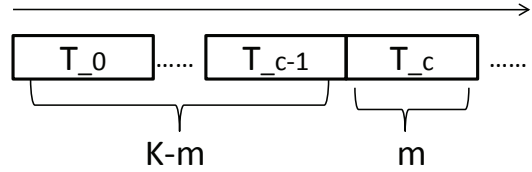}\\
  \caption{Dynamic Preservation Scheme based on Periodic Windows}
  \label{fig:slide}
\end{center}
\end{figure}

Recency requirement of TDIF application contains two aspects of demand.
The first is that users want to follow the recent information about a certain topic.
Secondly, for continuous data stream, the efficiency of information processing must be high.

Under the consideration of above two aspects, we propose a dynamic preservation scheme based on periodic time windows.
Specifically, we segment the online data stream into disjoint periods in time units (or in number of items).
Figure~\ref{fig:slide} is a simple example for illustration.
The core idea of scheme contains several aspects as following:
\begin{enumerate}
\item periodic time windows are \textit{disjoint} and \textit{non-overlapped};
\item utilizing the utility function as described in equation \ref{eq:comb}
\item utilizing \textit{reliant local preservation scheme.} Specifically, for each new time window, we preserve the $top\mbox{-}(k-m)$ items in prior result set, then utilize the utility function to select $m$ new items \textit{reliant} on the existing $k-m$ items. In this way, we can maintain \textit{diversity} of the final result set.
\end{enumerate}

The overall algorithm is described as Algorithm~\ref{alg:dtt}. When merging the old top-$(k\mbox{-}m)$ items and new $m$ items into result set, we strictly display the results in chronological order, which is described as line 6 in Algorithm~\ref{alg:dtt}. It can be described as the \textit{freshness} requirement of users in social media \cite{Drosou:2012}, where users are used to follow released information in chronological order.

The Algorithm~\ref{alg:dtt} is with time complexity of $O(|X^{(t)}|*m)$, $0<m \le K$, and $|X^{(t)}| \ll \sum_{t=0}^{T}{|X^{(t)}|}=N$.
Therefore, comparing with the traditional all batch mode which is with time complexity of $O(N*K)$, the dynamic mechanism will have better processing efficiency.
On the other hand, $m$ is a control parameter, which can flexibly control the ``staleness'' of the returned result set.
For example, if $m=K$, the Algorithm~\ref{alg:dtt} prefers to display the most recent information about the topic.

\begin{algorithm}[t]
\renewcommand{\algorithmicrequire}{	\textbf{Input:}}
\renewcommand{\algorithmicensure}{	\textbf{Output:}}
\caption {\textbf{Dynamic Preservation Scheme based on Periodic Windows}}
\label{alg:dtt}
\begin{algorithmic}[1]
\REQUIRE
$\textbf{S}_{K,t-1}~$- Result set with $K$ items until time $(t-1)$ \\
~~~~~~~~$X^{(t)}~$- The number of items in the new periodic time window
\ENSURE $\textbf{S}_{K,t}~$- Result set with $K$ items until time $t$
\STATE \text{Initialize:} $\textbf{S}_{K,t} \gets \text{top-(K-m) of~~} \textbf{S}_{K,t-1}$
\FOR{$i=1,...,m$}
\STATE $\text{bestDoc} \gets \mathop{\text{argmax}}_{x \in X_t} f_{S_{K,t}}(x,R)$
\STATE $\textbf{S}_{K,t} \gets \textbf{S}_{K,t} \cup \text{bestDoc}$
\ENDFOR
\STATE \text{Sort~}$\textbf{S}_{K,t}$ \text{by chronological order}
\STATE \textbf{return} $\textbf{S}_{K,t}$
\end{algorithmic}
\end{algorithm}

\section{Dynamic Diversity Evaluation Measures}

Topic-focused dynamic information filtering is a new application problem in current social media, which incorporates relevance, diversity, recency and confidence of information. Therefore, it is not easy to get a reasonable comprehensive evaluation for such a general task.

In the current Microblog task of TREC, the corresponding task evaluation only focuses on retrieval relevance \cite{mblog11,mblog12,mblog13}, and the detailed evaluation metrics are just the traditional MAP and P@K \cite{manning2008introduction}. While the diversity task of TREC Web track \cite{TREC09,TREC10,TREC11}, the corresponding evaluation metrics take both relevance and diversity into consideration, which contain $ERR\mbox{-}IA$, $\alpha\mbox{-}NDCG$ and $NRBP$. However, these existing measures can not take factors of recency and confidence into consideration, and are also not proper for the evaluation of TDIF application problem.
Based on the above analysis, we will propose a series of new dynamic diversity evaluation measures to get a more reasonable evaluation for TDIF task.

\begin{table*}[t]
\centering
\caption{Summary of typical diversity measures}
\label{tab:dcem}
\begin{tabular}{c|c|c|c|c} \hline
diversity&novelty&gain&discount&measure\\ \hline
\multicolumn{1}{c|}{\multirow{3}{*}{$\mathcal{S}=\frac{\sum_{i=1}^{M}p_i \mathcal{S}_i}{\mathcal{N}}$}}  & \multicolumn{1}{c|}{\multirow{3}{*}{$\mathcal{S}_i=\sum_{k=1}^{K}\frac{Q_i^k}{D_k}$}} &
\multicolumn{1}{c|}{\multirow{1}{*}{$Q_i^k = q_i^k \prod_{j=1}^{k-1}{(1-q_j^i)}$}} & $D_k=log(k+1)$ & $\alpha\mbox{-}NDCG$ \\
\cline{4-5}
 & & $\textit{or simplified to}$ & $D_k=k$ & $ERR\mbox{-}IA$ \\
\cline{4-5}
 & & $Q_i^k = g_i^k(1-\alpha)^{c_j^k}$ & $D_k=(1/ \beta)^{k-1}$ & $NRBP$  \\
\hline
\end{tabular}
\end{table*}

Firstly we will review the existing diversity evaluation measures that are summarized in table \ref{tab:dcem}.
These measures have the same nature, and are different in some tiny components such as the way of position discounting.
We find that there are 2 key points in these measures: \textit{diversity} and the \textit{gain}. The \textit{diversity} means subtopic (or aspect) coverage, which is based on explicit subtopic information of a query. Specific to a certain subtopic, the \textit{gain} describes redundancy penalizing and position discounting when accumulating the \textit{relevance} in every rank.
We take $\alpha\mbox{-}NDCG$ for example, $\alpha\mbox{-}NDCG$ is formulated as follows:
\begin{displaymath}
\alpha\mbox{-}NDCG = \frac{1}{\mathcal{N}}\sum_{i=1}^{M}p_i\sum_{k=1}^{K}\frac{g_i^k(1-\alpha)^{c_j^k}}{log_{2}(k+1)}
\end{displaymath}
where $g_i^k$ is a binary relevance value for document at postion $k$ with respect to subtopic $i$, $\alpha$ is a constant belong to $(0,1]$, $c_j^k=\sum_{j=1}^{k-1}g_i^j$, which is the number of documents ranked before position $k$ that are judged relevant to subtopic $i$, $K$ is the number of documents in a ranking list, $M$ is the number of subtopics, $p_i$ is the probability of each subtopic, and $\mathcal{N}$ is a normalization factor.

We incorporate recency and confidence factors into existing diversity evaluation measures such as $\alpha\mbox{-}NDCG$, and then propose a new dynamic diversity evaluation measure $d\mbox{-}NDCG$ as follows:
\begin{equation}
d\mbox{-}NDCG = \frac{1}{\mathcal{N}}\sum_{i=1}^{M}p_i\sum_{k=1}^{K}\frac{\gamma^ {t_{rcy}}*g_i^k(1-\alpha)^{c_j^k}*u_r}{log_{2}(k+1)}
\end{equation}
and $$t_{rcy}=topic.timestamp-tweet.timestap$$
where $topic.timestamp$ means the current time of topic tracking, $tweet.timestap$ means the released time of tweet information. $\gamma$ is the corresponding trade-off parameter, $0<\gamma \le 1$. we set $\gamma=0.5$ in our following experiment. $\gamma^ {t_{rcy}}$ part measures the recency of information.
$u_r$ measures the confidence of information via the way of user account weight \cite{Duan:2010:ESL,Chen:2011:TEI}.

Based on the definition of $d\mbox{-}NDCG$, we find that the final evaluation score of each items is depended on several factors: recency, relevance, diversity and confidence.
When in real application, we usually need to rescale the value of $t_{rcy}$ and $u_r$ upon the scale of relevance label $g_i^k$.
For example, the public twitter dataset in TREC Microblog task has three grade label: 2 (relevant), 1 (partly relevant) and 0 (not relevant). When in following experimental evaluation, we can simply rescale $t_{rcy}$ into three grade label: 2 (i.e.~history), 1(i.e.~recent) and 0(i.e.~latest) based on a certain threshold, and rescale $u_r$ into three grade label: 3 (i.e.~significant user account), 2 (i.e.~important user account) and 1 (i.e.~normal user account).

Similarly, we can give the corresponding definition of $d\mbox{-}ERR$ and $d\mbox{-}NRBP$, and simply replace the ``gain'' component in table~\ref{tab:dcem} with $\gamma^ {t_{rcy}}*g_i^k(1-\alpha)^{c_j^k}*u_r$, formalized as follows.
\begin{equation}
d\mbox{-}ERR = \frac{1}{\mathcal{N}}\sum_{i=1}^{M}p_i\sum_{k=1}^{K}\frac{\gamma^ {t_{rcy}}*g_i^k(1-\alpha)^{c_j^k}*u_r}{k}
\end{equation}

\begin{equation}
d\mbox{-}NRBP = \frac{1}{\mathcal{N}}\sum_{i=1}^{M}p_i\sum_{k=1}^{K}\frac{\gamma^ {t_{rcy}}*g_i^k(1-\alpha)^{c_j^k}*u_r}{(1/\beta)^{k-1}}
\end{equation}

\section{Experiments}

In this section, we will evaluate the TDIF task from different aspects. We first describe the experimental setup that includes dataset, evaluation metrics and baseline methods. Then we conduct extensive automatic evaluation for our approach and baseline strategies. Finally, we conduct manual evaluation for further analysis.

\subsection{Experimental Setup}

Here we give some introductions on the experimental setup, including data collections, evaluation metrics and baseline methods.

\subsubsection{Data Collections}

We use the public twitter dataset in Microblog task of TREC 2011 and TREC 2012, which has approximately a sample of 16M tweets, ranging over a period of 16 days. TREC 2011 provides 50 test topics, and TREC 2012 provides 60 test topics.

In our experiments, we only preserve English tweet data, and apply porter stemmer for tweet information and test topics.
Based on the consideration of ``short text'' of Microblog, we do not apply stopwords removing to avoid information loss.
We use Indri toolkit(version 5.2)\footnote{http://lemurproject.org/indri} as the basic retrieval platform. We also utilize the twitter API\footnote{https://github.com/lintool/twitter-tools} provided by TREC2013 to retrieval several features such as the number of followers and retweet number.
We conduct query expansion by pseudo relevance feedback and external expansion via Google search engine\footnote{http://google.com}, which aims to obtain more aspects of test topic for covering more information.

\subsubsection{Evaluation Metrics}

We will evaluate all the methods from two aspects of effectiveness and efficiency.
For effectiveness, we first utilize representative diversity measure $\alpha\mbox{-}NDCG$\cite{Clarke:2008}, and then utilize the proposed dynamic diversity measure $d\mbox{-}NDCG$. For $\alpha\mbox{-}NDCG$ and $d\mbox{-}NDCG$, the cutoff is set as $K=20$.

No matter $\alpha\mbox{-}NDCG$ or $d\mbox{-}NDCG$, they all need relevance label at subtopic level, while the current public dataset has not provided such information. Therefore, we do further manual relevance labeling at subtopic level, on the basis of existing all the relevant tweets. The labeling method is simple, for each relevant tweet, we judge whether it cover different subtopics comparing with prior relevant tweets. If yes, we will think it is relevant with a new subtopic.
We label 2955 relevant tweets for 49 test topics in total for TREC 2011, and label 6286 relevant tweets for 60 test topics in total for TREC 2012. On average there are 3.6 subtopics per test topic

For efficiency, we mainly utilize the average processing time of different methods for each test topic.

\subsubsection{Baseline Methods}

The R-LTR has been proved to be state-of-the-art diverse ranking methods. Therefore, in topic-focused information filtering task, we mainly focus on strategy comparison but not the detailed ranking models (or utility function). The typical baseline methods are shown as follows:
\begin{itemize}
\item All\_old. All\_old strategy means the original R-LTR method optimized for traditional diversity measures such as $\alpha\mbox{-}NDCG$, and then in each new time point, it will rank all the candidate items in a batch way.
\item All\_new. All\_new strategy denotes the R-LTR method optimized for new dynamic diversity measure such as $d\mbox{-}NDCG$, and rank all the candidate items in each time point.

\item TopRel. This method will select $K$ most relevant items in each new periodic time window. Specifically, it will use ListMLE method \cite{Xia:20086} as utility function, and display result in chronological order. This method does not consider the requirement of diversity, which is similar to the way used in industry.
\end{itemize}

Our proposed ``Dynamic reliant local Preservation scheme'' is denoted as ``DP'', which is based on the R-LTR utility function optimizing for $\alpha\mbox{-}NDCG$. If no special statement, the default value of parameter $m$ will be set as 10.

For proper evaluation, we choose `2 days' as a time unit, due to that there are not enough relevant tweets for each test topic in our dataset if we choose smaller time window size less than 2 days. Here we must state clearly that we can choose any proper window size based on the real application scenario.

We utilize the tweet data in first two days as training data, for utility function ListMLE and R-LTR, the detailed training process can be referred to the corresponding literature \cite{Xia:20086,zhu:2014:rRLTR}.

\subsection{Evaluation on Traditional Diversity Measure}

\begin{figure}[!htb]
\begin{center}
  \includegraphics[width=0.7 \textwidth]{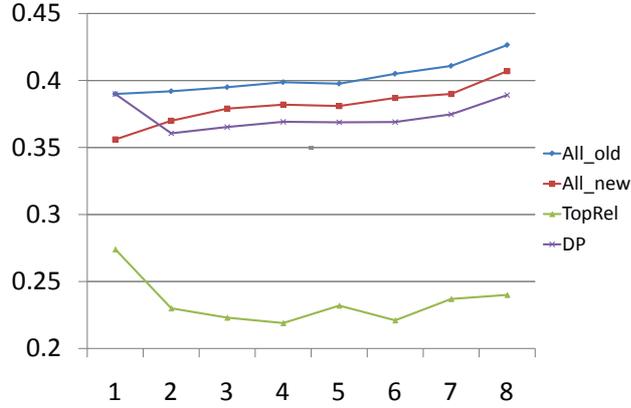}\\
  \caption{Performance comparison on $\alpha\mbox{-}NDCG$ measure}
  \label{fig:old}
\end{center}
\end{figure}

We first utilize traditional diversity measure $\alpha\mbox{-}NDCG$ for evaluation, and the detailed result is shown as figure~\ref{fig:old}. The horizontal axis means different time points in chronological order, and vertical axis denotes corresponding $\alpha\mbox{-}NDCG$ score.

From the figure, we can observe that All\_old performs best, which is in accordance with our intuition. All\_new also performs worse than All\_old due to optimizing for new diversity measure.
All\_Batch strategies (i.e., All\_old and All\_new) will rank all the candidate items in each time points. Therefore, they perform better than two other approaches.
Our DP approach shows less but approximate performance comparing with All\_Batch strategy. In fact, DP method can be viewed as an approximation of All\_old under online data stream scenario. It can capture more recency factors with the sacrifice of little performance on $\alpha\mbox{-}NDCG$.
TopRel performs worse because it only consider relevance requirement.
It can be applied easily, and used normally in industry filed.

\subsection{Evaluation on Dynamic Diversity Measure}

\begin{figure}[!htb]
\begin{center}
  \includegraphics[width=0.7 \textwidth]{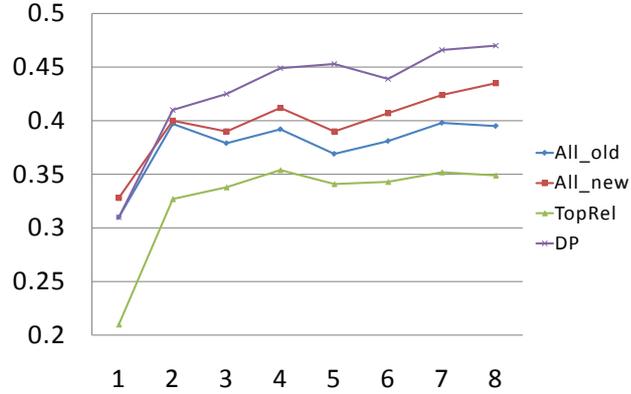}\\
  \caption{Performance comparison on $d\mbox{-}NDCG$ measure}
  \label{fig:new}
\end{center}
\end{figure}

The $d\mbox{-}NDCG$ is a new dynamic diversity measure, which also takes recency and confidence factors into consideration besides traditional relevance and diversity factors. Then we utilize $d\mbox{-}NDCG$ for further evaluation. The evaluation result is shown as figure \ref{fig:new}.

We can see that the proposed DP performs best among all baseline methods.
Although optimizing directly for $d\mbox{-}NDCG$ measure, All\_new still performs worse than DP strategy, which enforces capturing more recency factor based on time periodic window scheme.
Combing with the results in figure~\ref{fig:old}, All\_old and All\_new perform better under each optimizing diversity measure.
TopRel performs worst in all baselines, which is also consistent with the evaluation results in figure~\ref{fig:old}.

Overall, our proposed DP strategy shows better performance on $d\mbox{-}NDCG$ measure, which means our approach is more suitable for topic-focused dynamic information filtering task.

\subsection{Efficiency Evaluation}

\begin{figure}[!htb]
\begin{center}
  \includegraphics[width=0.7 \textwidth]{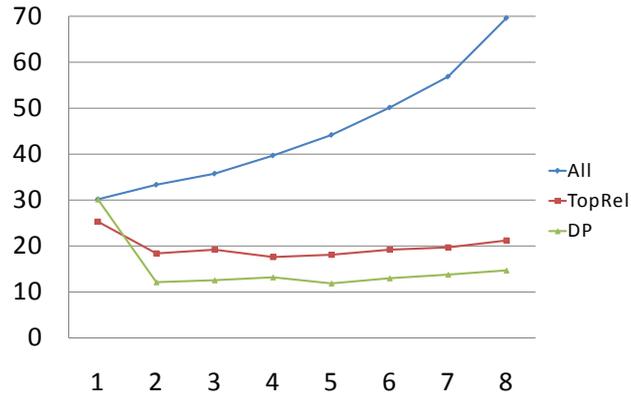}\\
  \caption{Average processing time of topic-focused information filtering (unit: millisecond)}
  \label{fig:effi}
\end{center}
\end{figure}

An important requirement of the TDIF task is the processing efficiency for online data stream.
Therefore, we will conduct efficiency evaluation with average processing time of each test topic.

The evaluation results are shown as figure~\ref{fig:effi}.
Here we use `All' denotes both All\_old and All\_new strategy since they are nearly with same efficiency.
We can see that All\_Batch strategy has lowest efficiency, because it will process \textit{all} the candidate items at each time point.
The DP strategy shows much higher efficiency than All\_Batch way, which is also consistent with the theoretical analysis in section 2.2. It will choose $m$ items in a candidate set with relatively small size at each time point.

TopRel shows lower efficiency than DP, but higher than All\_Batch.
Because it will choose 20 items in each time windows, and performs slower than Periodic approach (default $m=10$).
In fact, TopRel method drops the consideration of diversity relations, so it will perform faster than DP approach when $m=20$, which will be proved in the following evaluation of parameter $m$ sensitivity.

\subsection{Parameter Sensitivity}

\begin{figure*}[t]
 \centering
  \subfigure{
    \includegraphics[width=0.45 \textwidth]{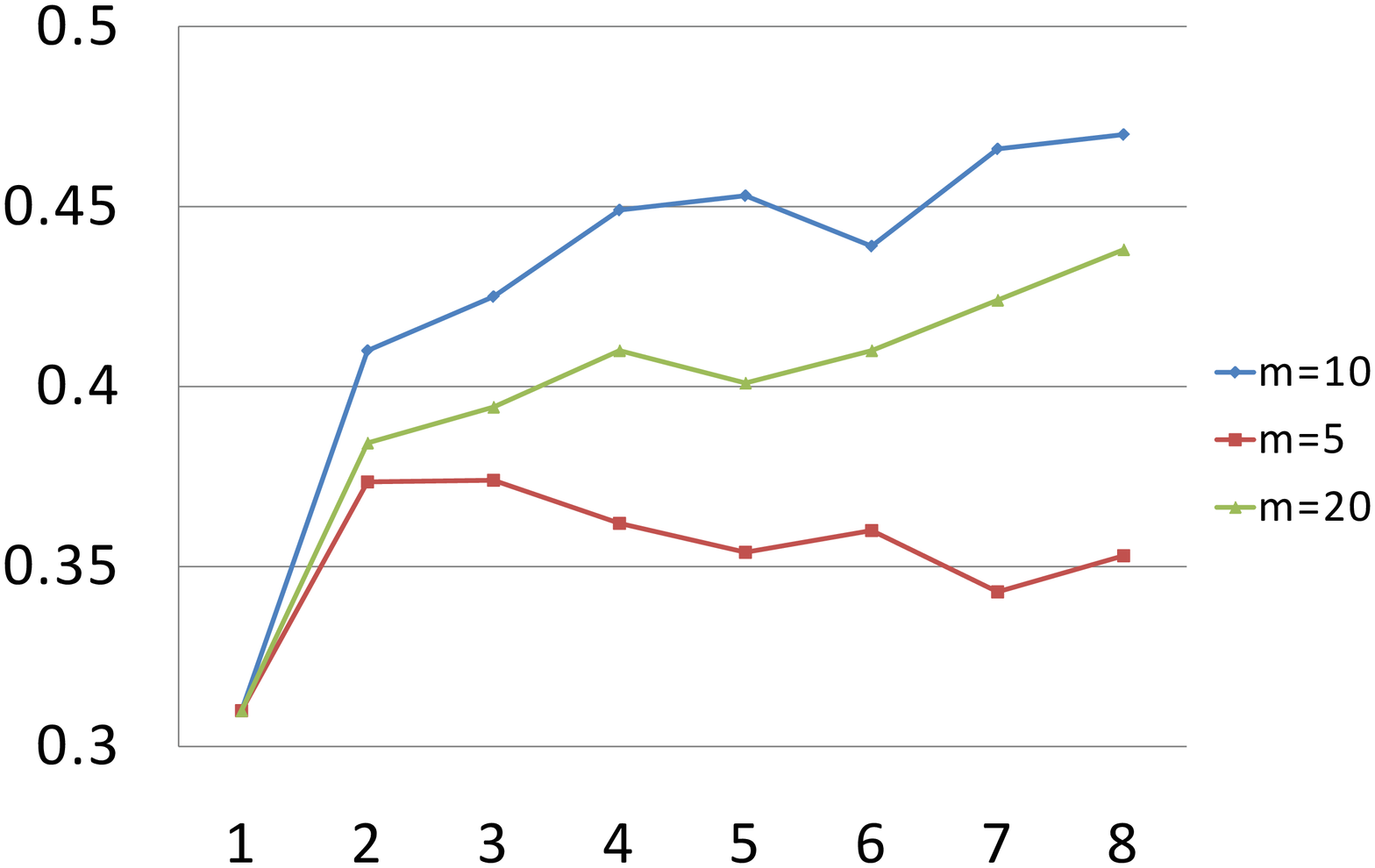}}
  \subfigure{
    \includegraphics[width=0.45 \textwidth]{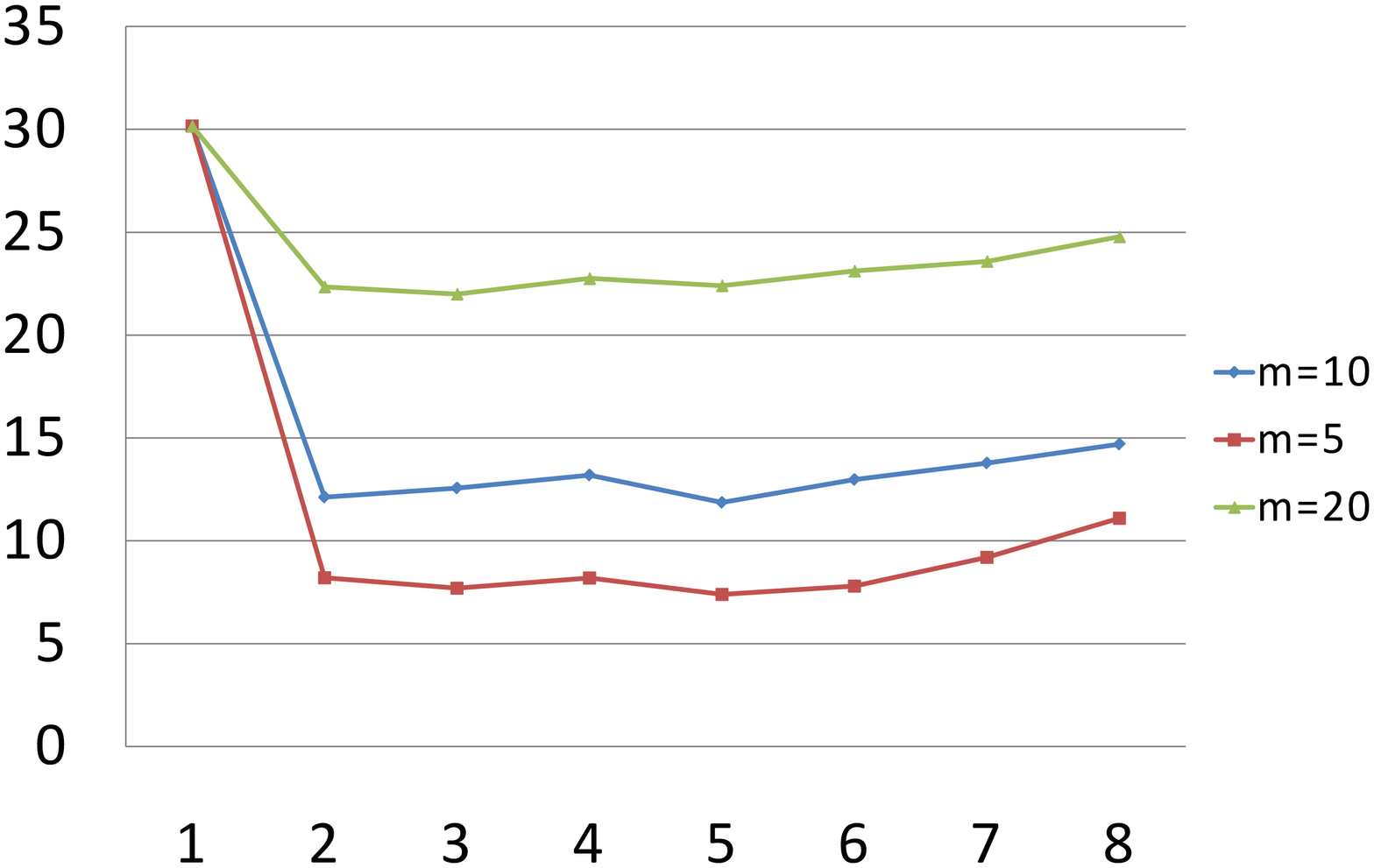}}
  \caption{Parameter sensitivity analysis of $m$: (a) Evaluation on $d\mbox{-}NDCG$ measure; (b) Average Processing time}
  \label{fig:sensi}
\end{figure*}

In our DP approach, the parameter $m$ ($0<m \le K$) control the ``staleness'' of the result set.
In this subsection, we will evaluate its effect from two aspects of $d\mbox{-}NDCG$ and efficiency.

We choose three situations of $m=5$, $m=10$ and $m=20$. The evaluation result is shown as figure~\ref{fig:sensi}.
From the performance of $d\mbox{-}NDCG$ (i.e.~subfigure~(a)), we can find that the case of $m=10$ performs best, and then followed with $m=20$ and $m=5$.
Form the aspect of efficiency (i.e.~subfigure~(b)), the case of $m=5$ performs best, and then followed with $m=10$ and $m=20$.
Therefore, based on the analysis of two aspects, $m=10$ will have better comprehensive performance, which is also set as default parameter value.

Additionally, when $m=20$, its processing time is during 20-25 milliseconds, which is slower than TopRel method (its average processing time is about 20 milliseconds, from figure~\ref{fig:effi}), due to the consideration of diversity relations.

\section{Related Work}

Most existing research work all treats the problem of diverse ranking as a `static subset problem' \cite{Agrawal:2009,Gollapudi:2009,Portfolio,Santos:2010,Rafiei:2010,proportionality}. Specifically, they will try to find optimal or suboptimal subset on a static data set. With the development of new social media such as twitter or sina weibo in china, the ranking scenario has changed. In this new scenario, new information will be continuously released online as a data stream, and how to process stream information effectively has become a new challenging problem.

The research work on the scenario of data stream is little, and several representative research work is \cite{DrosouP09,Minack:2011,Drosou:2012}. Drosou et al. \cite{DrosouP09} do some heuristic attempt on `` publish/subscribe'' scenario. Specifically, they give the definition of `diversity on sliding window', then utilize the classical ``Max-Sum'' object \cite{Gollapudi:2009} as utility function, to conduct heuristic greedy strategy. The idea of this work also inspires their following research work \cite{Drosou:2012}, which further focuses on the high efficient computing of dynamic diversity via an indexing scheme of ``cover tree''. It can support high efficient update operation such as inserting and deleting.
Mninack et al. give the definition of ``incremental diversity''. In their work, they can maintain a near optimal diverse set at any point in the data stream. The authors also utilize classical ``Max-Sum'' or ``Max-Min'' object as their utility function, to conduct heuristic interchange scheme. For each new items, it will make decision of discard or insert, to improve the diversity of the result set.

With the rise of social media, there are many related research work on social media. Chen et al. \cite{Chen:2011:SLW,Chen:2010:STE} discuss and analyze content recommendation in twitter from several feature dimensionality.
Hong et al. focus on how to build effective systems for ranking social updates from a unique perspective of LinkedIn. They leverage ideas from information retrieval and recommender systems, which has shown promising performance.
Choudhury et al. \cite{DeChoudhury:2011:IRS} focus on the research of topic retrieval in twitter, to obtain the most relevant results. However, their work is still limited to search scenario, which is almost same as traditional Web search.

Overall, comparing with prior research work, our work has shown several differences as follows:
(1)~the research problem is different, our work aims to tackle the topic-focused dynamic information filtering in social media, which is a new application problem; (2)~our detailed approach also shows many differences. We utilize different utility function - R-LTR ranking model, which is a supervised feature-based ranking model with good adaptation to different application scenario. Our dynamic preservation scheme also shows difference with prior work.

\section{Conclusions}

In this paper, we investigate the problem of topic-focused dynamic information filtering in social media.
Firstly we analyze the properties of the application problem, which has several typical requirements: relevance, diversity, recency and confidence.
In this scenario, how to balance these factors properly is very important.
Then we propose to utilize the relational learning-to-rank model, and combine with dynamic preservation scheme based on periodic time windows, to solve the TDIF problem. In this way, we can capture these ranking factors effectively.
Due to the new requirements of TDIF problem, we propose new dynamic diversity evaluation measures to get a more reasonable evaluation for such application problem, which can take recency and confidence factors into consideration on the basis of relevance and diversity.
We conduct extensive automatic and manual evaluation on public TREC twitter dataset, and the experimental results prove the effectiveness of our approach.

Overall, we present a completed investigation of a typical application problem in social media, which contains the analysis, solution and evaluation of the problem.
Our work shed some light on the TDIF problem, which is significative for future research work.

\scriptsize
\bibliographystyle{abbrv}
\bibliography{ctextemp_typeinst}

\end{document}